\documentclass{IEEEtran}

\usepackage{algorithm}
\usepackage{algorithmicx}
\usepackage[noend]{algpseudocode}
\usepackage{graphicx}
\usepackage{caption}
\usepackage{subcaption}
\newcommand{\correction}[1]{{{#1}}}
\usepackage{breakcites}
\usepackage{hyperref}
\usepackage{amsmath}

\begin{document}

\title{Improving Quality of Service for Users of Leaderless DAG-based Distributed Ledgers}

\author{Andrew Cullen\thanks{Andrew Cullen and Luigi Vigneri are with the Research Department at IOTA Foundation, Berlin, Germany.}, Lianna Zhao\thanks{Lianna Zhao and Robert Shorten are with Dyson School of Design Engineering at Imperial College London, London, United Kingdom.}, Luigi Vigneri, Robert Shorten\thanks{The first two authors contributed equally to this work.}}
\date{June 10, 2023}
\maketitle

\begin{abstract}
Usability of distributed ledgers is crucial to their mainstream adoption, especially for enterprise applications in which most users do not wish to operate full node infrastructure. Some attempts have been made to solve the problem of user-node interaction for blockchains in which leaders assemble users' transactions into blocks, but in the case of leaderless DAG-based ledgers such as IOTA, many of these solutions cannot be applied due to the absence of a shared mempool and the ability of nodes to issue blocks in parallel. In this work, we propose a user-node interaction mechanism for ledgers of this kind that is designed to balance user traffic across nodes and ensure the risk of a user experiencing a poor quality of service is low. Our mechanism involves users selecting nodes to issue their transactions to the ledger based on quality of service indicators advertised by the nodes. Simulation results are presented to illustrate the efficacy of the proposed policies. 
\end{abstract}

\section{Introduction}
The past decade has witnessed huge interest in the study of {\em distributed ledger technology} (DLT), both in the academic community \cite{rauchs2018distributed,seijas2016scripting,nguyen2018survey}, and in industry \cite{olnes2017blockchain,mills2016distributed,maull2017distributed}. The main topics of these studies on DLTs concern consensus mechanisms, scalability and security as well as a number of real use cases and applications, for instance in the areas of transportation \cite{zichichi2020distributed, vujivcic2020distributed, kamel2018geospatial}. However, as DLTs and their applications become better known and accepted in the wider technology community, the related issue of network usability and \emph{quality of service} (QoS) is being recognised as a bottleneck issue hindering applications of this nascent technology. Specifically, in many ledger architectures, users report lengthy and variable times for transactions to complete~\cite{zhang2020ethereum} and unpredictable fees~\cite{easley2019mining}. Furthermore, users of DLTs in the emerging realm of decentralised finance (DeFi) are subject to issues such as censorship and frontrunning at the hands of powerful block producers \cite{daian2020flash, kaihua2022quantifying}. As DLTs becomes more widespread and interoperable \cite{belchior2022do}, end users will choose the most reliable and user-friendly DLT available, but if distributed ledgers are ever to fulfil their true potential as viable decentralised alternatives to traditional systems, it is clear that QoS needs to be greatly improved---this is the focus of the work presented here. \newline

In many DLT architectures, the variation in the experience amongst users can be attributed to two factors. The first comes down to the design of core components of the DLT itself and how these affect the efficiency and performance of the system. For example, some DLT architectures have severe bottlenecks in their design that limit their ability to process new transactions efficiently. The second, equally important factor, is connected to the interaction between \emph{users} of the system and the \emph{nodes} that operate the DLT network, and the incentive structure that guides this interaction. This second factor is our primary concern here, as this has a strong influence on QoS for end users. To understand this factor more deeply, we need to consider the two kinds of actors involved, namely \emph{nodes} and \emph{users}. Nodes are individuals or organisations who maintain a copy of all transactions on the ledger and participate fully in network activities such as consensus and validation. Nodes have the power to modify the ledger to include transactions which they may use to transact themselves or may offer this as a service, perhaps for economic gain. For users (humans or software agents), on the other hand, the DLT is a tool to be utilised as part of some application, and their only objective is to send transactions through the network for a specific purpose. Users can only add data to the ledger by sending it to a node, usually via a digital wallet, and in many cases these users are willing to pay a fee to nodes to achieve a good QoS. The experience of users is strongly influenced by the manner in which {\em users} and {\em nodes} interact, which is driven by the different factors that motivate each actor's participation in a DLT network. Typically, {\em users} wish to transact as cheaply as possible while receiving some level of QoS from the network. For example, {\em users} may wish to transact with guaranteed instantaneous transaction delay, or with guaranteed maximum delay; minimize the financial cost of transactions or their energy consumption; or be simply assured of fairness and value for money they receive with respect to other users of the network and to be protected against exploitation and discrimination by nodes.{\em Nodes}, on the other hand, are driven by the desire to use of their resources (e.g., computational power or stake) for both personal or societal gain. \newline

The required mode of interaction between users and nodes depends heavily on the architecture of the underlying DLT, as does the core performance of the DLT, which influences the QoS that nodes can offer to users. In order to reduce delays and costs as well as preventing exploitation and censorship of users, we focus our attention in this work on DLTs based on a \emph{directed acyclic graph} (DAG) \cite{sompolinsky2016spectre, popov2018tangle} rather than the typical chain structure adopted by many market-leading DLT projects \cite{nakamoto2008bitcoin, buterin2014next, david2018ouroboros}. Our reasons for focussing on DAG-based ledgers are: their potential to support higher transaction throughput and lower delays due to their ability to process new data in parallel, meaning they can ultimately support better service for end users at lower costs; and their ability to support \emph{leaderless} consensus due to the possibility of adding multiple new blocks concurrently, which strips some of the power from block producers who otherwise would have complete power over what to include in their blocks. We will specifically focus on leaderless DAG-based DLTs such as \cite{muller2022tangle, baird2020hashgraph} for these reasons, and as we shall show, these architectures require a fundamentally different approach to improving QoS to end users than is required for their leader-based blockchain counterparts. \newline

\begin{figure}[h]
\centering
\includegraphics[width=0.9\columnwidth]{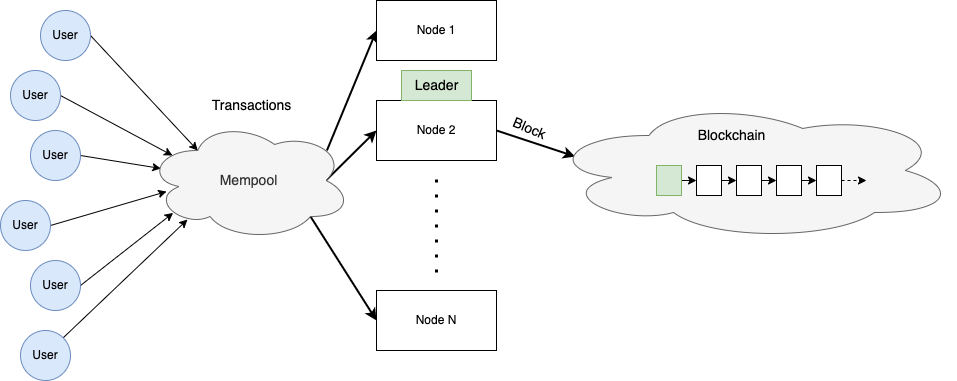}
\caption{The process for a user to write a transaction to a leader-based blockchain.}
\label{fig: blockchain process}
\end{figure}

\begin{figure}[h]
\centering
\includegraphics[width=0.9\columnwidth]{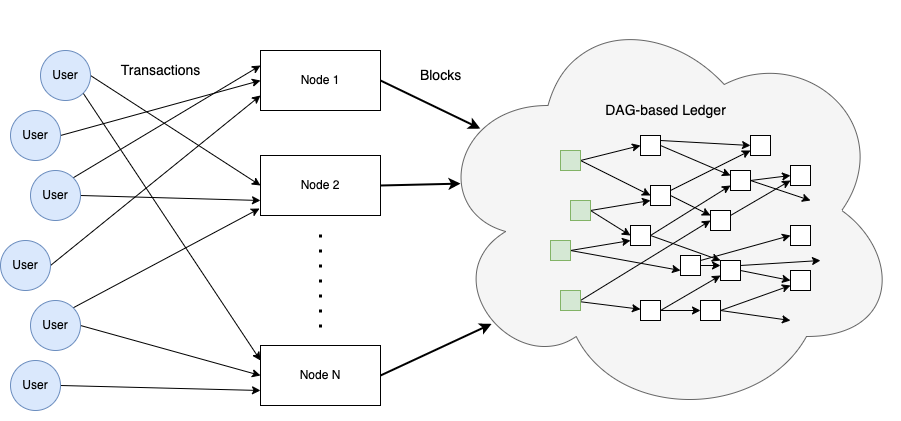}
\caption{The process for a user to write a transaction to a leaderless DAG-based DLT. }
\label{fig: dag process}
\end{figure}

To better understand the key differences between blockchains and leaderless DAG-based DLTs, let us outline the process required for a user to write a new transaction in each architecture. The process for a typical blockchain, as illustrated in Figure~\ref{fig: blockchain process}, proceeds as follows: the user shares the transaction to as many prospective block producers as possible, thereby adding their transaction to a shared \emph{mempool}, sometimes including a fee for the service; block producer nodes gather together transactions from the mempool into blocks, usually aiming to maximise their profits by including transactions with the highest fees; in each \emph{slot}, a single node is chosen to propose their block and write it to the ledger. The details of this process may vary slightly between blockchains, but the fundamental bottleneck always remains---all new transactions must be gathered up and placed in a block by a leader. It is this sequential block production bottleneck that not only limits transaction throughput but presents opportunities for block producers to exploit users, especially in the case of time-critical trading applications in the DeFi space \cite{zhou2022sok}. However, as we see in Figure~\ref{fig: dag process}, the process required in leaderless DAG-based DLTs has some crucial differences: users send each transaction directly to a chosen node rather than to a shared mempool; each transaction is placed in a block by the node that receives it and these blocks can be written in parallel to the ledger rather than sequentially by chosen leaders. \newline

\textbf{Remark:} the shared mempool model employed by blockchains is not suitable for leaderless DAG-based DLTs because data can written in parallel to the DAG. Hence, sending a transaction to multiple nodes would result in duplication of that transaction on the ledger. Duplication is not problematic for nodes, but is an inefficient use of the ledger. Furthermore, when contention for space on the ledger is high, this would be costly for the user who would have to pay a fee for each copy of their transaction.

Although the mode of interaction between users and nodes here differs significantly from a typical blockchain, the framework highlighted in Figure~\ref{fig: dag process} can be readily compared to infrastructures from other domains in which service users must choose from an array of service providers (nodes) and we can identify situations in which these interactions can lead to unstable network behaviour. For example, parallels can be found in road networks, where {\em toll roads} (nodes) are attractive to vehicles (users) precisely because they offer faster and more reliable transit times. However, if they attract too much traffic, precisely the opposite can be the effect, leading to unpredictable and sometimes catastrophic user experience. Problems of this nature are not unique to transportation and arise also in other domains - for example in job scheduling and queuing problems that arise in many applications (we shall have more to say on this in the next section when we present related work). Here we simply note that similar issues can arise in DLT networks if the interaction between users and nodes is poorly designed \cite{hunhevicz2020you} and while node-user interaction mechanisms have been designed for blockchains, there is currently no user-node interaction mechanism designed for DAG-based DLTs of the kind considered here. The problem of designing such a mechanism for DAG-based ledgers differs fundamentally from that of blockchains because nodes can add transactions to DAGs in parallel rather than sequentially as in blockchains. This feature of DAGs restricts users of these networks to selecting specific nodes to process each of their transactions and results in an entirely new paradigm for user-node interaction. \newline

Our goal in this paper is to suggest an interaction mechanism between nodes and users which allows both users and nodes to achieve their objectives and leads to networks that are stable, robust and fully utilised. Our idea is simple. Nodes broadcast a \emph{QoS indicator} for the service they can offer, and users respond probabilistically to this signal. We shall show that this simple algorithm equalises the QoS experienced by users, avoids large deviations in QoS experienced by individual users, whilst allowing nodes to be rewarded for the service they provide to the network. Moreover, the mechanism is inherently robust to the behaviour of dishonest nodes and users. Nodes advertising false QoS information can be rapidly detected and blacklisted by users. The proposed policies are also agnostic with respect to specific implementation issues, such as node discovery, and assume users have full access to QoS signals from nodes\footnote{In practice, nodes would most likely communicate their QoS signals to users by writing them to a commonly accessibly resource which can be easily queried by users as needed.}. As a final comment on the contribution of this work, it is important to note once again that the DAG-based DLT setting considered here requires a fundamentally different user-node interaction mechanism than those found in traditional blockchain architectures such as Bitcoin. The key difference lies in the fact that users must select a specific node to issue their transactions in our setting rather than broadcasting them to as many nodes as possible. To the best of our knowledge, this work is the first to consider the design of a user-node interaction mechanism for DLTs of this kind.
\newline 

The contributions of this work can be summarised as follows.
\begin{itemize}
\item{We introduce the problem of how to orchestrate the interaction of users and nodes in leaderless DAG-based DLT architectures for the first time and provide a model which captures the important nuances of the problem.}
\item{We propose a number of stochastic node selection policies to solve this problem which are inspired by research from other domains such as transportation networks and generalised job scheduling. These policies balance user traffic across nodes in a distributed manner, only requiring a QoS signal to be shared by nodes to users.}
\item{We present preliminary experimental results on a leaderless DAG-based DLT simulator which compare and contrast our stochastic node selection policies, ranging from naive policies to more sophisticated and effective ones.}
\end{itemize}

Our paper is structured as follows. In Section~\ref{sec: related work}, we present related work both from existing blockchain solutions and from other domains where similar problems arise. In Section~\ref{sec: model}, we give a basic system model for the relevant components of DAG-based DLT networks including some basic concepts, such as users, nodes and QoS indicators. In Section~\ref{sec: unim}, we present a user-node interaction mechanism for DAG-based DLT networks and propose a variety of stochastic policies for users and nodes to orchestrate their relationship. The proposed policies range from naive approaches to more sophisticated strategies which take into account different indicators of QoS. In Section~\ref{sec: sims}, a set of simulations are presented which validate and contrast the proposed policies. Finally, in Section~\ref{sec: conclusions}, we present out conclusions and propose some directions for future research.

\section{Related work}\label{sec: related work}
A wide variety of approaches have been taken to improving QoS for end users across various DLT architectures. One such approach is to settle payments \emph{off-chain} using tools such as lightning networks \cite{poon2016bitcoin}. Off-chain solutions of this kind allows users to make instantaneous payments on established channels without interacting with nodes. However, these solutions are limited to interactions between parties with some degree of trust established and can be thought of as side-stepping the real problem of improving QoS for users. 

Another way to avoid the complexities of users interacting with nodes is to purchase node infrastructure as a service. These service providers, inluding Infura\footnote{\url{https://www.infura.io/}} and Blockdaemon\footnote{\url{https://blockdaemon.com/}}, are primarily appealing to enterprise users that are willing to pay for the service. A similar path towards improved QoS can be achieved by enabling all users to act as block issuers themselves. To facilitate this, strenuous tasks required to issue a transaction were offered as a service by a third party provider in a early implementation of the IOTA network\footnote{https://ecosystem.iota.org/projects/powsrv-io}. The clear drawback of these approaches is that they rely on third party services, sacrificing some degree of decentralisation.

We are mostly interested in improving QoS by improving interactions between users and nodes. In leader-based blockchain architectures, the transaction fee mechanism plays a key role in determining how much users pay and how nodes decide how to prioritise transactions. For example, the first-price auction mechanism of Ethereum was a known issue inhibiting user experience with highly volatile fees and unpredictable service\footnote{\url{https://ethereum-magicians.org/t/eip-1559-fee-market-change-for-eth-1-0-chain/2783}}. Ethereum's primitive solution is EIP-1559 \cite{roughgarden2020transaction} which introduces a dynamic base-fee and an optional `tip', which is in line with the mechanisms employed by other market leaders such as Polkadot\footnote{\url{https://wiki.polkadot.network/docs/learn-transaction-fees}}. Transaction fee mechanism design in leader-based blockchains is discussed in \cite{roughgarden2021transaction}.

A crucial point in \cite{roughgarden2021transaction}, which partly motivates our interest in leaderless DAG-based DLTs, is that miners (nodes) in leader-based blockchains have dictatorial control over block contents, a fundamental property that can not be circumvented by any transaction fee mechanism. This can lead to issues with censorship as well as the emerging issue of \emph{maximal extractable value} (MEV) in which nodes manipulate transaction ordering in their blocks, often at the expense of users, to maximise their profits \cite{daian2020flash}. User exploitation of this kind is a relatively new problem but has already been studied extensively \cite{heimbach2022sok}. However, no prior research has studied MEV in the context of the leaderless DAG-based DLTs considered in this work, and while we expect that leaderless architectures may offer advantages in terms of MEV protection and censorship-resistance, we defer further exploration of this complex topic to future work.

The issue of user-network interaction also arises in several domains which are seemingly unrelated to DLT at first glance. For example, stochastic policies to guide this interaction (of this nature proposed here) have been studied and analysed in the context of transportation networks~\cite{crisostomi2020analytics, miller2000least, gehlot2020optimal, daganzo1977stochastic}. The authors in~\cite{caliskan2006decentralized} and further work in~\cite{idris2009car}  propose an algorithm enabling the number of arriving, departing cars and the instantaneous occupancy to be counted by car parks. Based on broadcasts from the car parks, cars then can predict the parking space available possibility at the estimated time that the car will arrive there. In~\cite{schlote2014delay} the authors propose a stochastic policy to associate cars with parking spaces and balance cars across a network of car parks and charge points, in a manner that minimizes the probability of a space not being available when cars arrive at a car part. The problem discussed in this paper is also closely related to a host of other load balancing problems that can be found in the networking literature; one example is that of server farms with immediate dispatching of jobs requiring task assignment policies \cite{baqer2016stressing}. Well-known job scheduling approaches that arise in this context include the {\em random selection policy}, {\em the round-robin selection policy} and the {\em join the shortest queue policy} \cite{semchedine2011task} (the algorithms proposed in the present paper are variants of the random selection policy in which a non-uniform distribution is sampled to select a node). Examples of relevant work in this direction can be found in~\cite{ciardo2001equiload, harchol1999choosing, schroeder2004evaluation, colajanni1998analysis, eager1986adaptive, ramamritham1989distributed} and the references therein. \newline

\section{System model}\label{sec: model}
\begin{figure}[H]
    \centering
    \includegraphics[width=0.85\columnwidth]{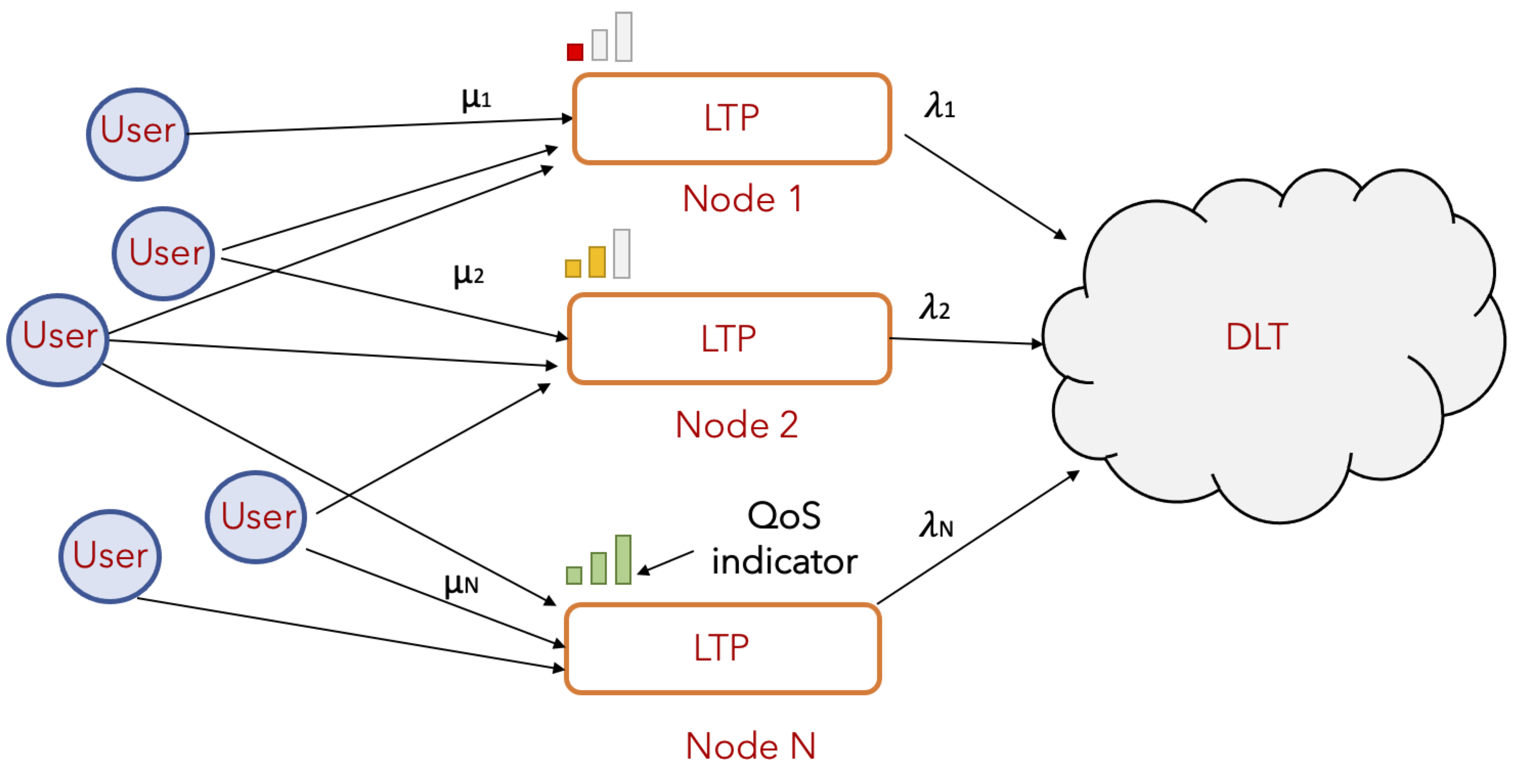}
    \caption{Basic model for user-node interaction in leaderless DAG-based DLTs. Transactions from users are added to the local transaction pool (LTP) of the receiving node.}
    \label{fig: Network}
    \end{figure}

We consider the network architecture as depicted in Figure \ref{fig: Network} which shows users accessing a DLT network via nodes, and nodes which communicate directly with one another, forming a distributed ledger network. More specifically, users make a selection from a set of nodes according to some criteria, and then send their transactions to these nodes to be processed. Furthermore, in this initial simplified model (neglecting some networking details such as availability of IP addresses) where users are able to query all nodes, users are free to select a node to process their transactions, and nodes are free to accept or reject transactions from individual users. It is important to note that, in contrast to blockchain-based DLT networks, users only send each transaction to a single node, and nodes do not share these transactions with one another prior to issuing them to the ledger. \newline

Nodes may offer ledger access as a public service and can offer different levels of QoS to users (some nodes may be operated by private organisations who reserve resources for their own gain, however, we do not consider such nodes any further in this work). Generally speaking, nodes must consume resources in order to add transactions to the ledger, and providing better QoS consumes more resources. For example, in a Proof-of-Work (PoW) ledger, the consumed resource is computation power, whereas in Proof-of-Stake (PoS) ledgers, the consumed resource is derived from each node's wealth in the native ledger currency. An extended version of PoS known as delegated PoS (dPoS) allows stake to be transferred to nominated nodes. Proof of Authority (PoA) \cite{de2018pbft} is an alternative approach for permissioned systems wherein each node's share of resources is configured from the outset of the DLT, for example, a consortium of companies could each be allocated an equal level of \emph{authority} in order to share control of a permissioned distributed ledger.
In the remainder of the paper, we refer to Proof of Reputation (PoR), which generalises PoS, dPoS, PoA and other resource allocation approaches. Reputation can refer directly to a node's stake (in the case of PoS), the stake delegated to it (in the case of dPoS), the authority assigned to it (in the case of PoA) or any other uninflatable quantity chosen for a given DLT. We denote the reputation of a node $i$ as $rep_i$. Although small inconsistencies in reputation calculation across nodes can typically be permitted and reputation can be changed over time, we assume that each node's reputation is a constant quantity that is agreed upon and publicly known. In what follows, we also make the following assumptions.

\begin{itemize}
\item Ledger access is limited by scarce resources so providing ledger access with low delay is costly to nodes. This is generally true for all DLT architectures (beyond the DAG-based DLTs considered here); for example, in a conventional PoW-based ledger, a node must consume energy at a greater rate to find PoW solutions more frequently and issue transactions with lower delay. \newline
\item Each node in our network is equipped with a Local Transaction Pool (LTP) that is used as a buffer to store pending transactions. Transactions sent from a user to a particular node enter that node's LTP. Note that similar pools of transactions can also be found in other ledger architectures, such as the mempool in the Bitcoin network, although nodes typically share a common mempool which is not permitted in our setting. \newline
\item Nodes can advertise their expected QoS that they can offer via some proxy.  For example, nodes may advertise the expected delay from receiving a new transaction to them writing to the ledger. Additionally, nodes may require a fee for issuing a transaction and this can also be taken into account as part of a node's overall QoS. \newline
\end{itemize}

A model for our user-node interaction mechanism is depicted in Figure \ref{fig: Network}. The sending rate of user $i$ is denoted by  $\mu_i$, while the service rate\footnote{Service rate refers to the rate at which nodes can successfully write transactions from their LTP to the ledger.} of node $i$'s LTP is denoted by $\lambda_i$. The QoS indicator for each node is represented by the colored bars above their LTP. Here, a red QoS indicator simply means poor QoS is offered by this node (for example, this could be due to high delay and/or high fees), while yellow indicates average QoS and green indicates good QoS. 

In what follows, we are specifically interested in designing a mechanism to orchestrate the interaction between users and nodes in leaderless DAG-based DLT networks. Our goal is to deliver a uniformly good QoS to all users and ensure that the probability of any user experiencing a very bad QoS is low, while allowing nodes to profit by processing a regular stream and (on average) fair share of transactions. While our results generally apply to any leaderless DAG-based DLT, we include some details of IOTA's implementation as a concrete example. Specifically, the rate at which nodes can issue transactions in the IOTA network is regulated by an access control algorithm~\cite{cullen2021access, zhao2021secure}. Each node regulates its own issue rate via a distributed algorithm based on the additive increase multiplicative decrease (AIMD) algorithm~\cite{cai2005performance, shorten2006positive, shorten2005analysis}: nodes increase their issuing rate additively until congestion is detected locally, at which point they decrease their issuing rate multiplicatively which allows each node to converge to an issuing rate proportional to their reputation. By locally following this AIMD algorithm, each node obtains an issuing rates proportional to its \emph{reputation} (for example, in a PoS system, they would achieve an issuing rate proportional to their stake), and these issuing rates are then enforced throughout the network via a scheduling algorithm based on deficit round robin (DRR) algorithm \cite{shreedhar1995efficient}. \newline 

While the definition of QoS is quite broad, we will focus our attention mostly on delay and fees as measures of QoS in this work. In order to provide an indicator of QoS to users, nodes must estimate the expected delay for them to issue a newly arriving transaction. This delay depends on two factors: the number of transactions in the node's LTP at the time of the new transaction's arrival; and the service rate of the LTP. As mentioned above, in the IOTA network, the service rate of each node's LTP is determined by the AIMD-based rate setting algorithm employed by the node (see the Appendix for further detail). Given this background, we now propose policies by which users and nodes can interact with each other in order to achieve good network behaviour.

 \section{User-Node Interaction Mechanism}\label{sec: unim}
We now propose simple but powerful policies by which users and nodes can interact with each other in order to achieve good network behaviour. Specifically, we propose stochatic policies for users with the objective of minimising their probability of experiencing bad QoS when connecting to a given node. These stochastic policies perform the role of a load balancer, but operate in an entirely distributed manner with no central agent required to route traffic. Additionally, we propose a policy for nodes to adapt their fees and incentivise users to reduce their demand in the presence of congestion. Our policies generally operate as follows.
\begin{itemize}
\item Nodes measure the expected delay that is associated with processing a transaction. Nodes may additionally require a fee for their service that can be adapted in response to congestion, and this can also be provided as a QoS indicator for users.
    
\item Users gather these QoS indicators provided by the nodes and construct a vector $p$ representing a probability distribution over the nodes. Users then sample this distribution to select a node based on a probability that is proportional to the QoS indicator (inversely proportional to expected delay/fees) of a given node.
\end{itemize}

\begin{figure}[H]
    \centering
    \includegraphics[width=0.98\columnwidth]{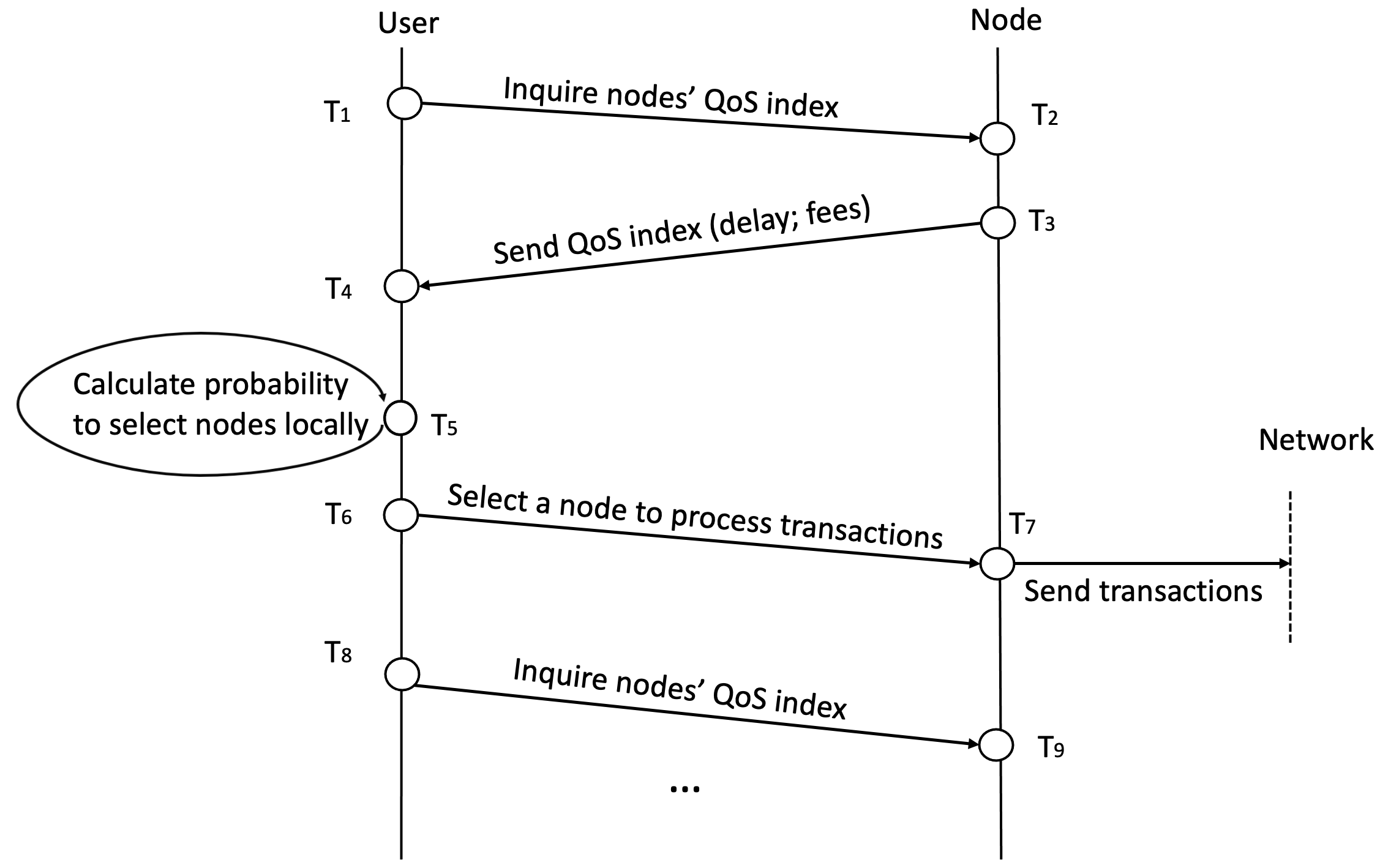}
    \caption{User-node interaction mechanism.}
    \label{fig: Node_user}
    \end{figure}

In Figure~\ref{fig: Node_user}, we show an example of this user-node interaction in which users directly query nodes for their QoS indicators. In practice, nodes would most likely broadcast these values by writing to a commonly accessible resource for the users to query which could provide a more efficient solution.

Stochastic policies of this nature have been studied and analysed in the context of transportation networks \cite{miller2000least, gehlot2020optimal,daganzo1977stochastic} and we shall not repeat this analysis here. Intuitively, these policies can be thought of as implementing a distributed load balancer---when all users operate this policy, nodes are kept busy, and users minimize their probability of experiencing a large delay without the need for any third party or centralised load balancer.

While such policies operate well in transportation and mobility networks, a defining characteristic of DLT based environments is that they are adversarial, so the security of our proposed protocol must be taken into account. Our mechanism can be adapted to deal with an adversarial setting in a number of ways. For example, \emph{spamming attacks} in which users send large amounts of traffic to nodes, can be dealt with by requiring users to complete a small PoW or by introducing fees as we shall demonstrate in Section \ref{subsec: fees}. An additional attack vector for users could be to create double spends and send each spend to a different node to issue: this kind of attack can be easily dealt with by nodes by requiring users to make a verifiably random selection using a public source of randomness such as a distributed random number generator \cite{syta2017scalable} and hashing this with some unique identifier of the user to produce their random node selection.

Additionally, there may be users not following the suggested protocol, and this can not be determined definitively until implemented in a real system with human actors involved. However, the choice faced by users can be intuitively thought of in a similar manner to the choice faced by network users in TCP---following the stochastic policy is not a Nash equilibrium because an individual user could benefit from deviating from this policy, but if all users defected then they would all be worse off. For example, if a user chose to simply deterministically send their transaction to the node with highest QoS, they may get good service, but if many nodes did this, the highest QoS node would become overwhelmed and all these users' service would suffer. Just as in TCP, if the user's choice can be abstracted away effectively and the default option is set to our stochastic policy, it is reasonable to expect that most users will not deviate from this.\newline

In the algorithms outlined in this section, $\mathcal{N}$ is the set of all available nodes, $p$ is a vector whose $i^\text{th}$ element, $p_i$, represents the probability of selecting node $i$, and $rep_i$ is node $i$'s reputation. In this work, we assume that $\mathcal{N}$ includes all nodes, i.e., all nodes make their service available to all users. In reality, some nodes may not wish to make their services available to users or may even make them available to a select group of users, perhaps on a subscription basis. However, we defer further considerations of such complexities to future work.


\subsection{Naive policies}
To provide a benchmark for comparison to the policies we will propose, we begin by describing two stochastic policies that may seem reasonable. We refer to these as {\em naive} policies as they do not attempt in any serious manner to shape traffic reaching the nodes based on prevailing network conditions. It is important to reiterate that no state of the art exists in the context of DAG-based DLTs, as user-node interaction usually refers to blockchain where a common mempool is used. The first policy we describe is a simple \emph{uniform random node selection} (URNS) policy, as outlined in Algorithm \ref{alg: urns}. The second, less naive, policy which we will compare to our proposed policies is referred to as \emph{reputation-based node selection} (RBNS), as described in Algorithm \ref{alg: rbns}. \newline

\begin{algorithm}[h]
    \caption{URNS executed by users}  \label{alg: urns}      
    \begin{algorithmic}[1]
    \State $p_{j} \leftarrow 1/|\mathcal{N}|, \forall j \in \mathcal{N}$ \Comment{Initialise probabilities} \newline
    \Statex \emph{Repeat each time a transaction is sent:}
    \State \hskip1em $j \gets$ random sample from $p$
    \State \hskip1em Send transactions to node $j$
    \end{algorithmic}
\end{algorithm}
    
\begin{algorithm}[h]
    \caption{RBNS executed by users}  \label{alg: rbns}      
    \begin{algorithmic}[1]
    \State $p_{j}=\frac{rep_j}{\sum_{i \in \mathcal{N}} rep_i}, \forall j \in \mathcal{N}$ \Comment{Initialise probablities} \newline
    \Statex \emph{Repeat each time a transaction is sent:}
    \State \hskip1em $j \leftarrow$random sample from $p$
    \State \hskip1em Send transactions to node $j$
    \end{algorithmic}
\end{algorithm}
As can be seen from the above algorithms, URNS involves selecting nodes uniformly at random, without taking in to account any information about the node whatsoever. RBNS, on the other hand, makes use of the globally known reputation of each node which gives some information about the rate at which each node can issue new transactions. Both URNS and RBNS are \emph{open loop} policies in the sense that they do not make use of feedback from nodes about the current QoS indicators such as delay as we do in the policies which we propose next. As we shall see in Section \ref{sec: sims}, closed loop policies based on measured QoS metrics offer significant improvement over these open loop alternatives.


\subsection{Selection policy based on pure delay}
In order to improve upon the naive policies stated above, we would like to introduce QoS metrics from nodes. We first focus our attention on delay as the primary indicator of QoS. As such, we refer to our solution as \emph{delay-based node selection} (DBNS) as presented in Algorithm \ref{alg: dbns}.

\begin{algorithm}[htp] 
\caption{DBNS executed by users} \label{alg: dbns}
\begin{algorithmic}[1]

\Statex \emph{Repeat each time a transaction is sent:}
\State \hskip1em
            $p_{j} \gets \frac{rep_j/\tau_j}{\sum_{i \in \mathcal{N}} rep_i/\tau_i}, \forall j \in \mathcal{N}$ \label{line: prob1} \Comment{Update probabilities}
\State \hskip1em $j \gets$ random sample from $p$
\State \hskip1em Send transaction to node $j$
\end{algorithmic}
\end{algorithm}

Algorithm \ref{alg: dbns} is executed by a user each time it wishes to have a transaction issued to the ledger. The probability update of line \ref{line: prob1} is the critical step of the algorithm, namely:
\begin{equation*}
p_{j}=\frac{rep_j/\tau_j}{\sum_{i \in \mathcal{N}} rep_i/\tau_i}
\end{equation*}
where $p_{j}$ is the probability that a user choose node j, $\tau_i$ is the expected delay in node $j$'s LTP, and  $rep_i$ is the reputation of node $i$. Note that this policy is decentralised, requiring no information about the operation of other users. The only information required by users from each node is their reputation and their expected delay. Moreover, recall that the reputation of each node is a quantity that is known by all users. \newline

The expected delay, on the other hand, can be calculated by nodes at regular intervals and broadcast to all users. The delay signal, $\tau_i$, broadcast by nodes acts as a feedback to users about the state of the network, resulting in a closed loop system. The expected delay for a transaction arriving to node $i$'s LTP at time $t$ can be estimated as follows:
\begin{equation}
\tau_i(t) = \frac{L_i(t)}{\tilde{\lambda_i}(t)} \label{con:Estdelay}
\end{equation}
where $L_i(t)$ is the number of transactions in node $i$'s LTP at time $t$ (including the newly arrived transaction). $\tilde{\lambda_i}(t)$ is the expected average service rate of the LTP over the time this new transaction spends waiting to be issued. In a PoW ledger, this would be a fixed value, $\tilde{\lambda_i}(t)=\lambda_i$, based on the computing power of node $i$ (provided the node's computing power remains constant). In this work, however, we focus our attention on the IOTA protocol in which the service rate of the LTP is governed by an AIMD rate setter \cite{cullen2021access} as part of IOTA's access control algorithm which replaces the need for PoW. We employ a max-min moving average filter to estimate the issue rate of each node as described in the \emph{Appendix}.

\subsection{Including transaction fees} \label{subsec: fees}

The above policy can be extended to include transaction fees which offers improved security for nodes against malicious users and demonstrates the ability of our approach to capture broader notions of QoS. This extension requires a policy for nodes as well as for users. Specifically, nodes require a fee $\sigma_i$ to be included in any transaction and they broadcast this along with their expected delay $\tau_i$, to users. When congestion is detected by nodes, they increase their fee to reduce the load on them. To achieve this goal, a simple P controller (it also can be adapted to PI/PID equivalent) is adopted here for node $i$.
\begin{equation} \label{eq: fee}
\sigma_i(t) = \max{\left(0, K_{pi} (\tau_i(t) - r_i)\right)}
\end{equation}
where $K_{pi}$ is the proportional gain (a positive constant) and $r_i$ is a delay setpoint. If the expected delay of node $i$'s LTP, $\tau_i(t)$, is less that $r_i$, then node $i$ will set their fee to zero, and as delay increases beyond this setpoint, the fee $\sigma_i(t)$ will increase at a rate proportional to $K_{pi}$. Both $K_{pi}$ and $r_i$ can be tuned by nodes on an individual basis, but we will assume that they are equal for all nodes in this work. \newline

Users, on the other hand, can compute a cost function based on their desired trade-off between fee and delay to get a QoS metric for each node. The cost function for user $m$ can be specified as follows
$$c_{mi}(t) = a_m \tau_i(t) + (1-a_m) \sigma_i(t)$$
where $c_{mi}(t)$ is the QoS metric used by user $m$ for node $i$ at time $t$, $a_m$ determines the weights assigned to delay $\tau_i(t)$ and fees $\sigma_i(t)$ for user $m$. Larger values for $a_m$ indicate that user $m$ cares greatly about having low delay whilst lower values reflect that this user is more concerned about paying low fees.
User also define a cost threshold, $c_m^{max}$, above which this user will not continue sending transactions, which means when congestion increases, fees go up and some users stop sending transactions until the QoS metric is lower than their threshold. The algorithm employed by users, which we call DBNS+, including both delay and fees, is described in Algorithm \ref{alg: dbns+}.

\begin{algorithm}[htp] 
\caption{DBNS+ executed by user $m$} \label{alg: dbns+}
\begin{algorithmic}[1]

\Statex \emph{Repeat each time a transaction is sent:}
\State $\mathcal{N}^* \gets \mathcal{N}$
    \For{$j \in \mathcal{N}$}
    \State $c_{mj} = a_m \tau_i + (1-a_m) \sigma_i$
        \If {$c_{mj} > c_m^{\mathrm{max}}$}
    \State remove $j$ from $\mathcal{N}^*$
    \EndIf
\EndFor
\If{$\mathcal{N}^*$ is not empty}
    \State $p_{j} \gets \frac{rep_j/c_{mj}}{\sum_{i \in \mathcal{N}^*} rep_i/c_{mi}}, \forall j \in \mathcal{N}^*$ \label{line: prob2}
    \State $j \gets$ random sample from $p$
    \State Send transaction to node $j$
\EndIf
\end{algorithmic}
\end{algorithm}


 \section{Simulations}\label{sec: sims}
\correction{
The simulation results presented in this section are produced with a Python simulator originally designed for testing IOTA's DAG-based distributed ledger, in which multiple nodes communicate with one another over simulated communication channels and each maintains a copy of the ledger. The simulator was modified specifically for this work to include a local transaction pool at each node and generate traffic from a large number of external simulated users  \cite{simulator}. In our experiments, 50 nodes are available to issue transactions for users, where each node in the network is peered with 4 randomly chosen neighbours such that the communication graph of the nodes has a random 4-regular topology. The mean propagation delay of each communication channel between neighbours is chosen uniformly at random between 50 ms and 150 ms, and the delay for each block on these channels is normally distributed around this average with standard deviation 20 ms. No bootstrapping procedure is required for nodes in this simulation because all nodes are present from the outset of the network, which simplifies the simulations. The nodes we discuss in this paper are obeying restrictions imposed by the access control algorithm in \cite{zhao2021secure} wherein the issuing rates of nodes are controlled by an AIMD algorithm which allows them to increase their issuing rates when the network is uncongested and then decrease their rate when congestion occurs. \newline

Transactions issued by nodes need to be forwarded to neighbours to achieve a shared view of the ledger, and the total rate at which nodes can process new transactions is given by a parameter, $\nu$. In the below simulations, $\nu$ is set to 50 transactions per second. All simulations here are run for a duration of 3 minutes of simulation time in steps of 10 milliseconds, which is sufficient to observe the stationary properties of the system because no bootstrapping of the network is required. Results are averaged over 10 Monte Carlo simulations, unless otherwise specified. The reputation distribution of the nodes follows a Zipf distribution with exponent 0.9, as depicted in Figure \ref{fig:RepDist}. These statistics are chosen to be representative of real values that have been measured from the IOTA network\footnote{This model is well suited to reputation systems derived from wealth, i.e., PoS\cite{jones2015pareto}, and we can observe distributions of this kind in wallet balances in IOTA \url{https://thetangle.org/address/richlist-and-distribution}.}.
 \begin{figure}[H]
    \centering
    \includegraphics[width=0.8\columnwidth]{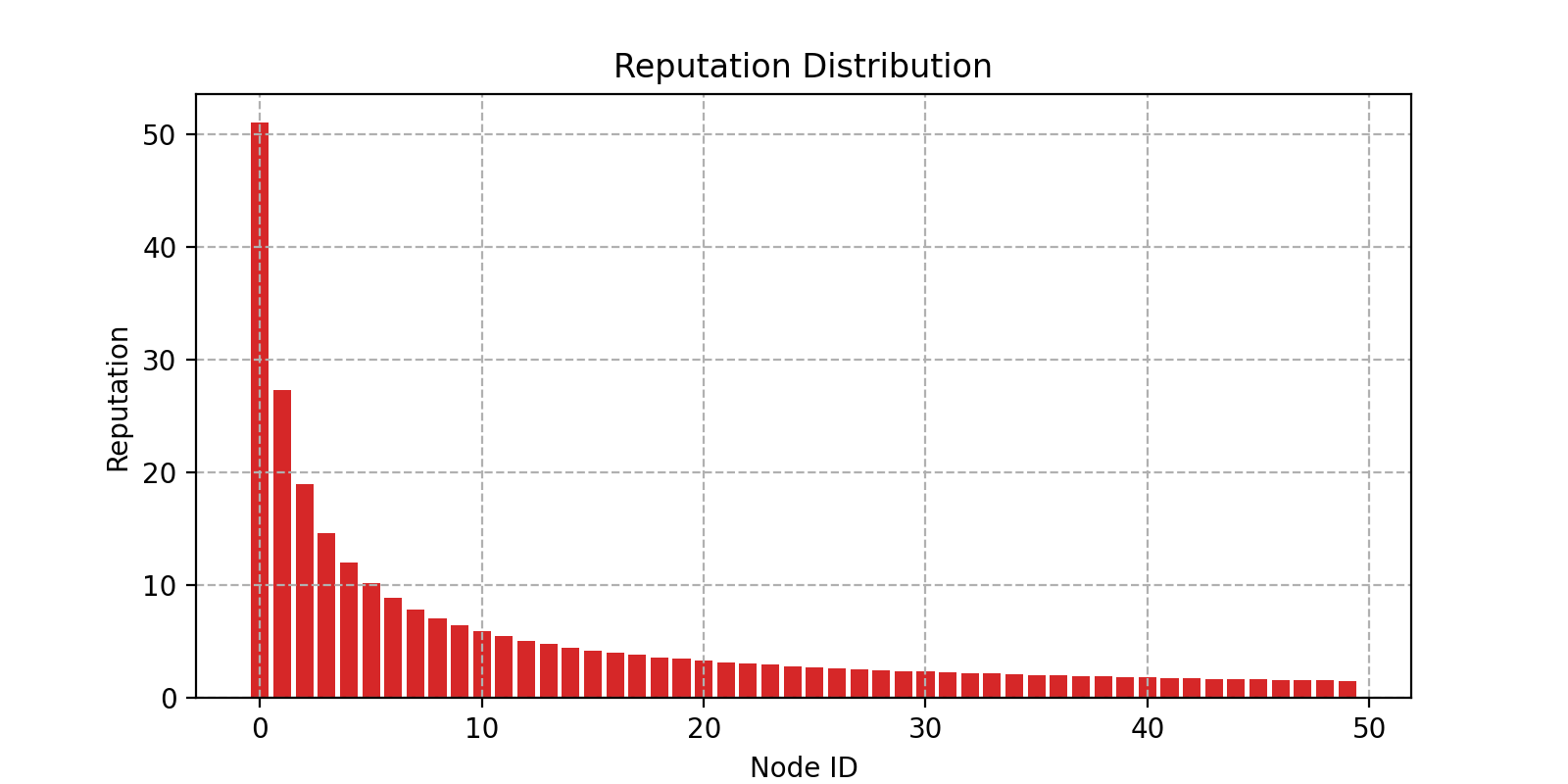}
    \caption{Reputation distribution of nodes in all simulations.}
    \label{fig:RepDist}
 \end{figure}
 
Note that URNS, RBNS and DBNS (Algorithms~\ref{alg: urns}--\ref{alg: dbns}) do not require any special parameters to be specified, but for simulations in which DBNS+ (Algorithm~\ref{alg: dbns+}) is employed, the following parameters are used.
 \begin{itemize}
 \item $K_{pi}$, the proportional gain is set to $0.8$ for all nodes.
 \item $ r_i(t)$, the desired level of delay is set to $15$ seconds. 
 \item $a_m$, which specifies the weighting of delay and fees to each user is set to $0.6$ for all users.
 \item $c_m^{\mathrm{max}}$, the cost threshold for users is set randomly between $8$ and $10$ for each user to capture some variation in what users expect in terms of delay and fees.
 \end{itemize}
All the above parameters are chosen experimentally for the purpose of the preliminary results we present in this paper, and more extensive simulations with a range of parameters should be the subject of future research.

We organise our simulation results as follows. Experiments were conducted by changing the node selection policy employed by users from the portfolio of stochastic policies:
\begin{itemize}
\item[(i)] URNS which selects nodes uniformly at random (see Algorithm~\ref{alg: urns});
\item[(ii)] RBNS which selects nodes based on their reputation (see Algorithm~\ref{alg: rbns});
\item[(iii)] DBNS which selects nodes based on their reputation and delay (see Algorithm~\ref{alg: dbns});
\item[(iv)] DBNS+ which builds on DBNS by including transaction fees (see Algorithm~\ref{alg: dbns+}). 
\end{itemize}
Apart from changing the node selection policy employed by users, all other simulation parameters are the same in each scenario.

For each group of simulations (i)--(iv), we consider three different scenarios based on the combined sending rate of users which represents the total traffic which nodes must process. In all scenarios, each new transaction is generated by a new user, and these transactions follow a Poisson arrival process. Poisson processes are standard for modelling events from a large population of independent agents. Each new user is independent of all others and they select a node according to the policy being tested. Each policy is tested in isolation, i.e., we do not consider mixtures of different policies from different users in this work. The three scenarios are as follows, according to the average sending rate of users as a percentage of the total network scheduling rate.
\begin{itemize}
\item[(a)] 90\%---a scenario where demand is below capacity, although transactions do not arrive uniformly over time (they follow a Poisson arrival process), so the instantaneous arrival rate will sometimes exceed the scheduling rate of the network.

\item[(b)] 98\%---a scenario which demonstrates behaviour very close to full capacity.

\item[(c)] 120\%---a scenario which simulates how these approaches deal with congestion in high-load situations above the network capacity.
\end{itemize}

  \begin{figure*}[t!]
  \centering
  \begin{subfigure}[t]{0.31\linewidth}
  \includegraphics[width=\linewidth]{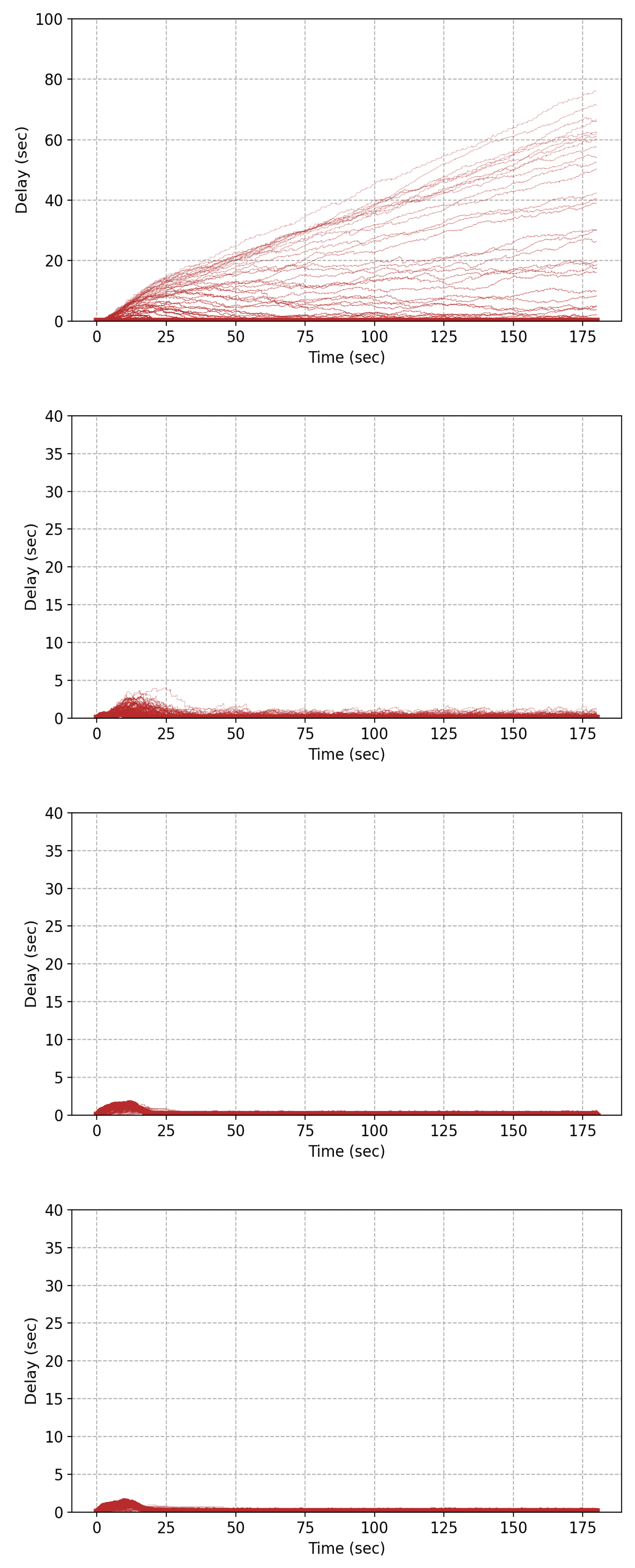}
  \caption{90\% capacity.}
  \label{fig:delay0.9}
  \end{subfigure}
  \hspace{1mm}
  \begin{subfigure}[t]{0.31\linewidth}
  \includegraphics[width=\linewidth]{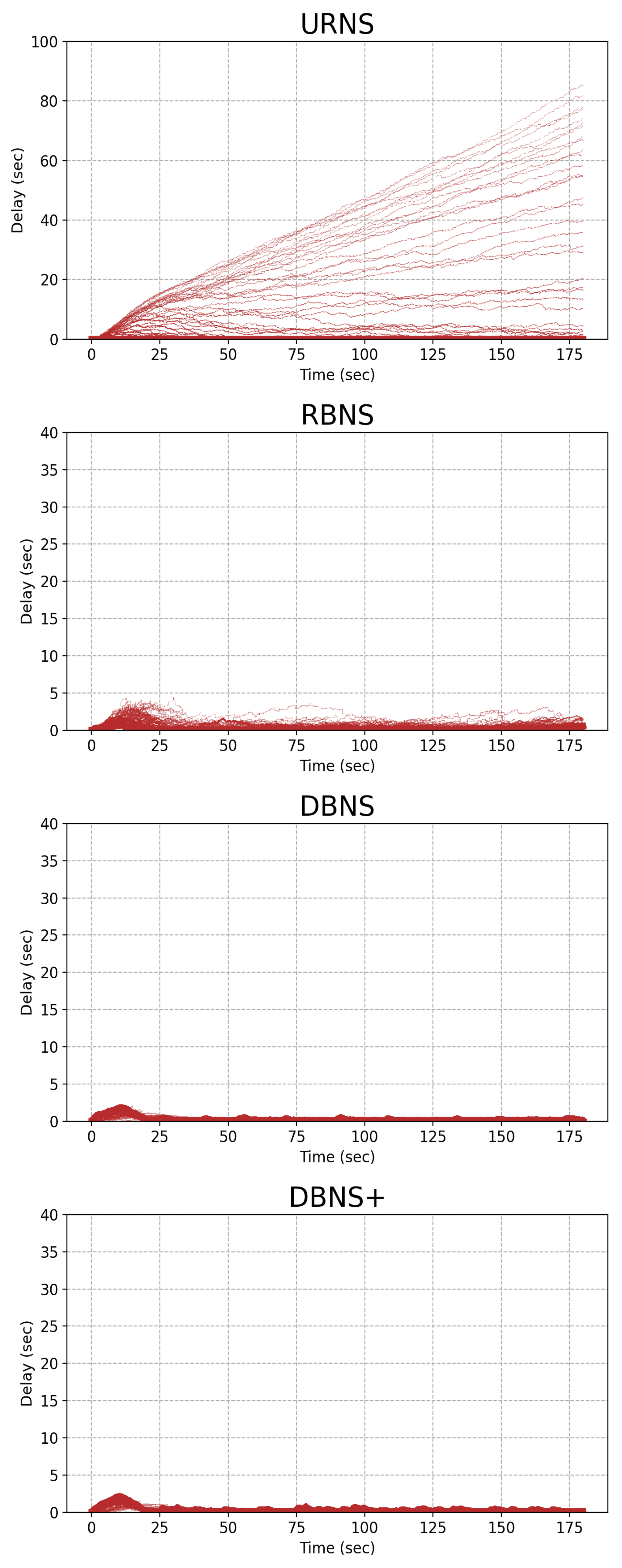}
  \caption{98\% capacity.}
  \label{fig:delay0.98}
  \end{subfigure}
  \hspace{1mm}
  \begin{subfigure}[t]{0.31\linewidth}
  \includegraphics[width=\linewidth]{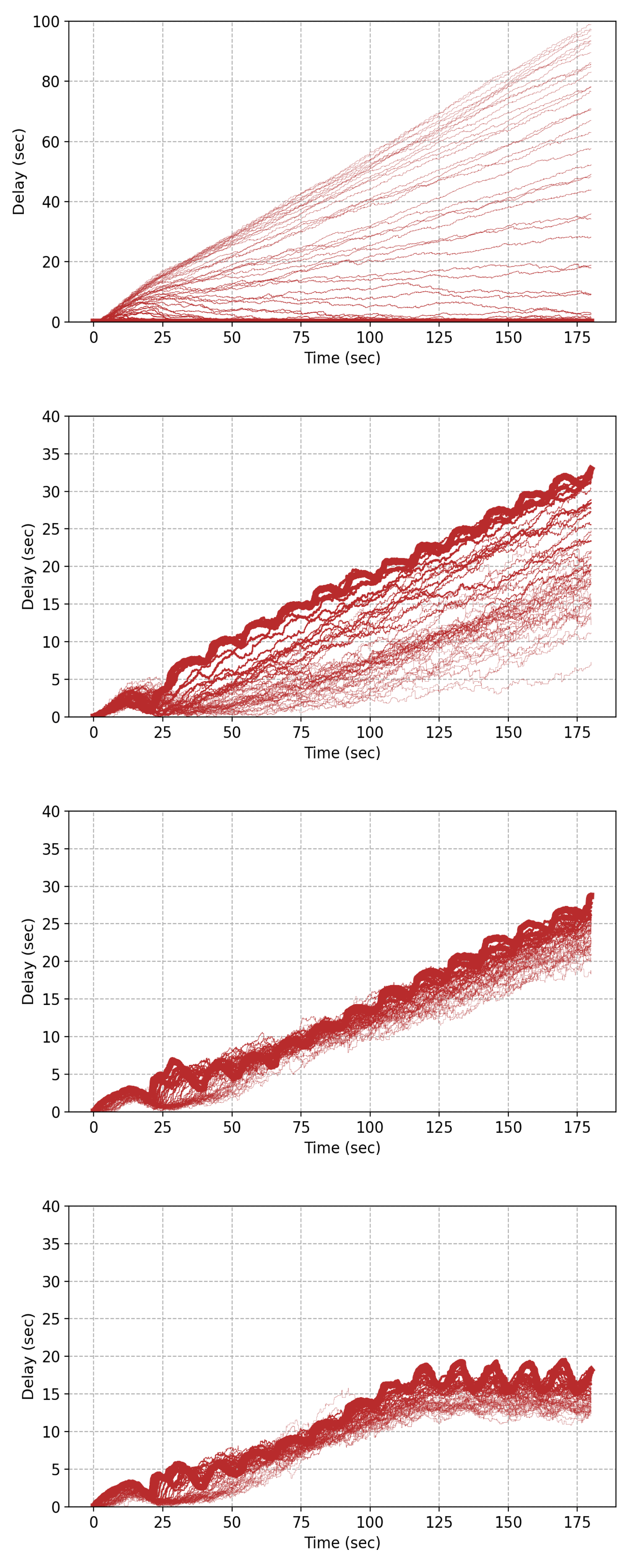}
  \caption{120\% capacity.}
  \label{fig:delay1.2}
  \end{subfigure}
  \caption{Measured transaction delay in each node's LTP under URNS, RBNS, DBNS and DBNS+ policies.}
  \label{fig: delays}
  \end{figure*}
\subsection{Load Balancing}
We begin by examining how well each policy balances user traffic across the 50 nodes in our experimental setup. This can be effectively shown by plotting the delays experienced in each node's LTP as in Figure~\ref{fig: delays}. The results are arranged in a grid where each row corresponds to one of the policies (i)--(iv), and the columns correspond to a traffic scenario (a)--(c). Within each plot, we show the delay in each node's LTP, where the thickness of each trace is proportional to the reputation of the node. 
\subsubsection*{(i) URNS policy}
For each scenario, (a)--(c), the first row of Figure~\ref{fig: delays} plots traces for the delays experienced at each of the 50 nodes when users employ the URNS policy. We observe that lower reputation nodes experience severe delays and high reputation nodes are underutilised and hence experience very low delays. From the perspective of nodes, traffic from users is very poorly balanced, resulting in low reputation nodes being overwhelmed and high reputation nodes being underutilised. We can conclude that URNS performs poorly across all scenarios. \textbf{Note:} the y-axis scale on the first row differs from the other three approaches due to the relatively poor performance of URNS compared to the other three less naive policies.\newline
  
\subsubsection*{(ii) RBNS policy}
The second row of Figure~\ref{fig: delays}, which shows the LTP delays of nodes in the RBNS simulations, the improvement over the URNS policy is immediately evident. We can see from these plots that in scenario (b), at 98\% capacity, fluctuations in delay of some nodes become more evident due to the fact that no feedback is used here. In the high load scenario of 120\% capacity, we see that higher reputation nodes experience higher delays which is related to the fact that the issue rates of nodes are not perfectly fair with respect to their reputation. It is clear that a closed loop approach is required to achieve better fairness. \newline

\subsubsection*{(iii) DBNS policy}
The third row of Figure~\ref{fig: delays} shows results for the DBNS policy. We see that there is less fluctuation in scenario (a) and (b) than was the case for RBNS, and in case (c), we also observe that the traffic is balanced more fairly among nodes. These improvements over RBNS can be attributed to the use of feedback in DBNS (users base their choice of Node the congestion level in their LTP as well as their reputation). However, there is still no policy in place to control congestion in the LTP effectively with this policy, so DBNS still performs poorly in scenario (c) with delay continuing to grow throughout the simulation.\newline
  
\subsubsection*{(iv) DBNS+ policy}
Finally, we present preliminary results for DBNS+. As can be seen from the bottom row of Figure~\ref{fig: delays}, in scenarios (a) and (b), DBNS+ performs similarly to DBNS, but when we introduce more severe congestion in scenario (c), there are clear advantages to the DBNS+ policy. In particular, we see in scenario (c) that the delay of each node is effectively stabilised by DBNS+. The reason for the stabilisation of delays after a certain point is that nodes increase their fees in response to congestion which increases the QoS metric, $c_{mi}(t)$, for users. As a result, some users will eventually exceeds the threshold $c_m^{\mathrm{max}}$. At this point, users stop sending transactions to node $i$. Note that the oscillations evident in the plots for scenario (c) are a direct result of the AIMD algorithm governing the service rate of the LTP which in turn determines the estimated delay serving as feedback to users.

\subsection{User Experience}
The results provided by Figure~\ref{fig: delays} primarily focus on the perspective of nodes rather than directly considering the experience of users. To capture the user perspective more effectively, we first consider the statistics provided by Table~\ref{tab: expected ltp delay} which gives the expected delay experiences by a user.

\begin{table}[h]
\caption{Expected LTP delay}
\centering
\begin{tabular}{r | c | c | c |}
 & (a) 90\% capacity & (b) 98\% capacity & (c) 120\% capacity \\
\hline
URNS & 10.83 s & 12.84 s & 17.24 s \\
\hline
RBNS  & 0.29 s & 0.40 s & 11.80 s \\
\hline
DBNS  & 0.15 s & 0.26 s & 11.58 s \\
\hline
DBNS+  & 0.15 s & 0.21 s & 10.17 s \\
\hline
\end{tabular}
\label{tab: expected ltp delay}
\end{table}

For the purpose of contrasting policies with respect to these user experience statistics, we compare the policies in each traffic scenario rather than considering each policy in turn, as we did for the load balancing results.

\subsubsection*{(a) 90\% capacity} 
The first column of Table~\ref{tab: expected ltp delay} compares the four policies for scenario (a). URNS stands out for its exceptionally poor performance, with an expected LTP delay for users of 10.83 seconds. The other three policies offer significant improvements with 0.29 seconds for RBNS and 0.15 seconds for both DBNS and DBNS+.
\subsubsection*{(b) 98\% capacity} 
The second column of Table~\ref{tab: expected ltp delay} shows similar results for scenario (b) in which we increase the load relative to scenario (a). URNS leads to an even higher expected LTP delay of 12.84 seconds, while delays remain low for the other three approaches, each policy steadily improving on the last. 
\subsubsection*{(c)120\% capacity} 
In the high load scenario (c), we only see a slight improvement in each more advanced policy, with the expected delay of DBNS+ just over a 1 second less than that of DBNS. However, we expect that if these simulation were run for longer, we would see a larger difference due to the growing LTP delays in DBNS as we observed in the third row of Figure~\ref{fig: delays}. \newline

The above results only show the expected LTP delay, but when it comes to user experience, some users may not care about the exact delay of their transaction as long as it does not exceed some threshold that they deem unreasonable. To take this aspect of user experience into account, consider Table~\ref{tab: qos20}, which provides the probability that a user experiences a delay of greater than 20 seconds. The threshold of 20 seconds applies well to the scenarios considered here but a different acceptable threshold may be considered depending on the needs of users in a given application. This statistic can be thought of as the probability that a random user will get unlucky and have a bad experience, which may turn them off using the technology again in the future.
\begin{table}[h]
\caption{Probability of LTP delay greater than 20 seconds.}
\centering
\begin{tabular}{r | c | c | c |}
& (a) 90\% capacity & (b) 98\% capacity & (c) 120\% capacity \\
\hline
URNS & 0.21 & 0.25 & 0.32 \\
\hline
RBNS  & 0.00 & 0.00 &  0.25 \\
\hline
DBNS  & 0.00 & 0.00 & 0.20 \\
\hline
DBNS+  & 0.00 & 0.00 & 0.01 \\
\hline
\end{tabular}
\label{tab: qos20}
\end{table}

\subsubsection*{(a) 90\% capacity} 
The first column of Table~\ref{tab: qos20} once again shows us that URNS performs worst of all the policies. In the case of this metric, we see that there is a probability of 0.21 that a randomly chosen user in this experiment will experience a delay greater than 20 seconds if all users employ URNS, while the other three policies ensure that delays greater than 20 seconds never occur.
\subsubsection*{(b) 98\% capacity} 
Similarly, the second column of Table~\ref{tab: qos20} shows that when traffic is at 98\% of capacity, there is a probability of 0.25 that a randomly chosen user will experience a delay greater than 20 seconds with URNS, while the probability is 0 for all other policies. 
\subsubsection*{(c)120\% capacity} 
The most interesting results for this metric can be found in the third column of Table~\ref{tab: qos20} which shows the results for each policy with demand from users at 120\% of capacity. In this case, we see that URNS, RBNS and DBNS all perform poorly, with probability of a randomly chosen user experiencing delay greater than 20 seconds of 0.32, 0.25 and 0.20, respectively. However, this is where we see the improvement offered by DBNS+, for which there is only a probability of 0.01 of delay exceeding 20 seconds. It is worth noting once again that the threshold of 20 seconds is selected here because it applies well to the scenarios and parameters chosen, but some threshold related to a meaningful delay should be chosen when evaluating these policies in practice.
}
\section{Conclusion}\label{sec: conclusions}
In this paper, we proposed a user-node interaction mechanism for DAG-based DLTs which seeks to improve the usability of such networks for basic users who do not wish to run a fully operation node. The mechanism involves nodes calculating QoS metrics and providing this information to users. Users then use this feedback in a stochastic policy to select a node. Experiments including contrast experiments were carried out to validate the effectiveness of the proposed algorithms. In particular, our experiments show that by combining fees and measurements of expected delays, QoS can be provided in a fair manner to users and the demand from users can be allocated fairly among nodes, and that each of our proposed policies improves upon the last in terms of performance. However, each policy also increases in complexity, and this is worth taking in to consideration when implementing these policies in practice. For example, RBNS is shown to provide reasonably good performance when demand is slightly lower and nodes are not congested, and it does so without requiring any QoS signals, so this may be a sufficient choice in some cases.
 There are a number of simplifying assumptions used in this work, for example, we assumed that users have full knowledge of nodes statistics, which is probably unfeasible for large networks. For future work, it would be interesting to study how our policies perform when users query only a subset of nodes for QoS indicators. It would also be interesting to consider heterogenous user behaviour and to introduce more complex node behaviours such as collaboration between nodes and sharing of fees through mechanisms such as Shapley value. Other work to be considered include attack analyses for the policies presented here.

\bibliographystyle{IEEEtran}
\bibliography{refs}

\appendix 
\section*{appendix}\label{appendix}
\begin{figure}[H]
\centering
\includegraphics[width=0.8\columnwidth]{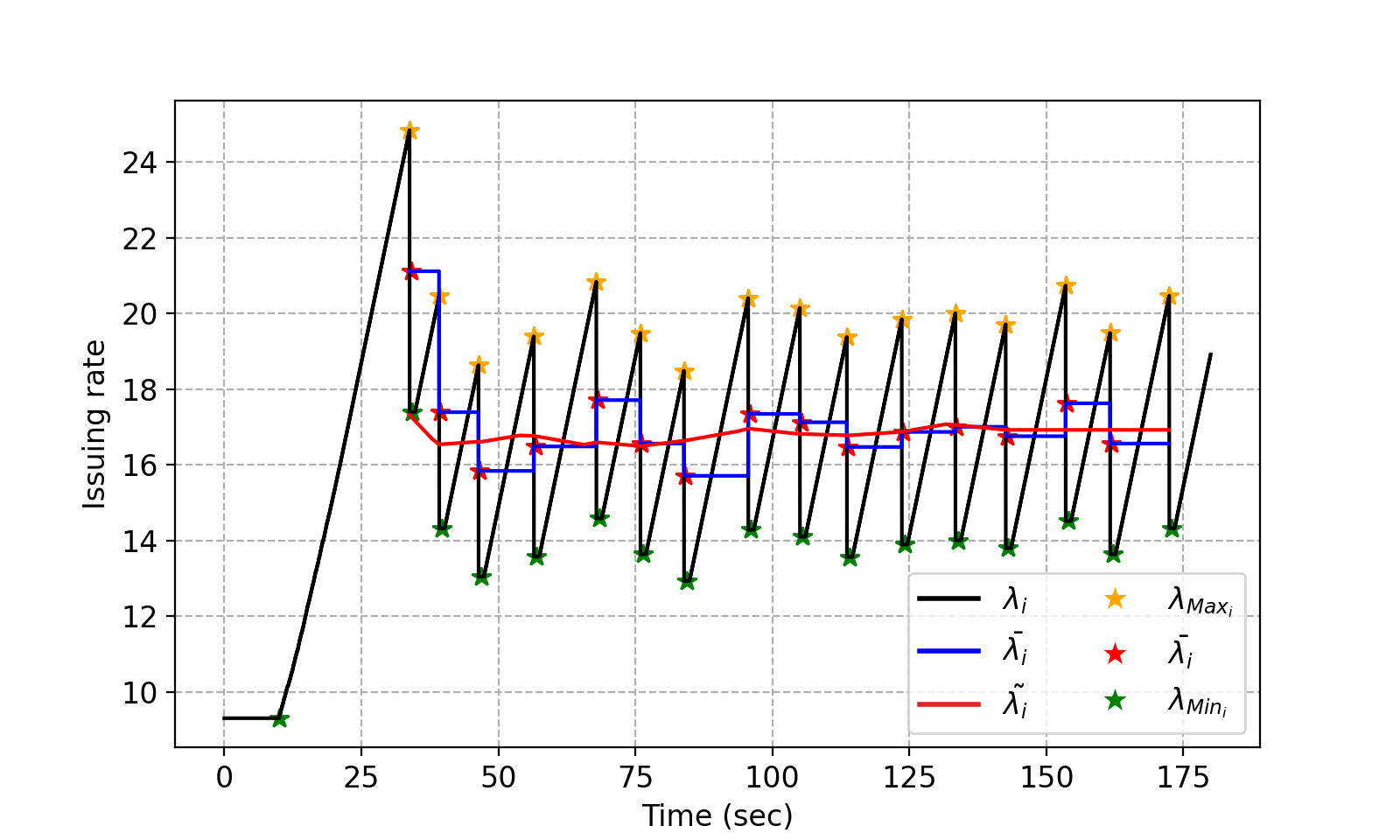}
\caption{An example of applying max-min moving average filter for one node. }
\label{fig: max_min_filter2}
\end{figure}

In our simulations nodes use a max-min moving averaging filter. This filter is based heavily on knowledge of the IOTA access control protocol; in particular, on the fact that IOTA uses an AIMD algorithm as part of the rate scheduling algorithm. A typical node-issuing-rate, based on the AIMD algorithm, is depicted in Figure~\ref{fig: max_min_filter2}. As can be seen from the plot, the issuing rate varies over sawtooth shaped cycles in which the issuing rate moves from a minimum to a maximum issuing rate according to the AIMD algorithm. Based on this knowledge the filter operates as follows. 

\begin{itemize}
\item Each time a minimum of the sawtooth is detected, a node records the corresponding issuing rate. We denote this value $\lambda_{Min{_i}}(k)$ where the index $k$ refers to the $k^{th}$ cycle.\newline 
\item When the next maximum of the sawtooth is detected, a node records this value, denoted $\lambda_{Max{_i}}(k)$.\newline  
\item The node then averages the minimum and maximum rate over the $k^{th}$ cycle. We denote this quantity by $\bar{\lambda_i}(k)$.\newline
\item Finally, the node averages these values,  $\bar{\lambda_i}(k)$, $\bar{\lambda_i}(k-1)$,....,$\bar{\lambda_i}(k-\delta)$, to obtain a filtered estimate of the issuing rate delay which is named $\tilde{\lambda_i}(k)$.\newline 
\end{itemize}
Finally, an estimate of the delay for each transaction after the $k^{th}$ cycle is calculated as in~\eqref{con:Estdelay}.

\end{document}